\documentclass[ALICE,manyauthors]{cernphprep}
\usepackage[comma,square,numbers,sort&compress]{natbib}
\usepackage{hyperref}
\usepackage{lineno}
\usepackage{xspace}
\usepackage[T1]{fontenc}
\usepackage{orcidlink}
\begin{document}
%

\newcommand{\pp}           {pp\xspace}
\newcommand{\ppbar}        {\mbox{$\mathrm {p\overline{p}}$}\xspace}
\newcommand{\XeXe}         {\mbox{Xe--Xe}\xspace}
\newcommand{\PbPb}         {\mbox{Pb--Pb}\xspace}
\newcommand{\pA}           {\mbox{pA}\xspace}
\newcommand{\pPb}          {\mbox{p--Pb}\xspace}
\newcommand{\AuAu}         {\mbox{Au--Au}\xspace}
\newcommand{\dAu}          {\mbox{d--Au}\xspace}

\newcommand{\s}            {\ensuremath{\sqrt{s}}\xspace}
\newcommand{\snn}          {\ensuremath{\sqrt{s_{\mathrm{NN}}}}\xspace}
\newcommand{\pt}           {\ensuremath{p_{\rm T}}\xspace}
\newcommand{\meanpt}       {$\langle p_{\mathrm{T}}\rangle$\xspace}
\newcommand{\ycms}         {\ensuremath{y_{\rm CMS}}\xspace}
\newcommand{\ylab}         {\ensuremath{y_{\rm lab}}\xspace}
\newcommand{\etarange}[1]  {\mbox{$\left | \eta \right |~<~#1$}}
\newcommand{\yrange}[1]    {\mbox{$\left | y \right |~<~#1$}}
\newcommand{\dndy}         {\ensuremath{\mathrm{d}N_\mathrm{ch}/\mathrm{d}y}\xspace}
\newcommand{\dndeta}       {\ensuremath{\mathrm{d}N_\mathrm{ch}/\mathrm{d}\eta}\xspace}
\newcommand{\avdndeta}     {\ensuremath{\langle\dndeta\rangle}\xspace}
\newcommand{\dNdy}         {\ensuremath{\mathrm{d}N_\mathrm{ch}/\mathrm{d}y}\xspace}
\newcommand{\Npart}        {\ensuremath{N_\mathrm{part}}\xspace}
\newcommand{\Ncoll}        {\ensuremath{N_\mathrm{coll}}\xspace}
\newcommand{\dEdx}         {\ensuremath{\textrm{d}E/\textrm{d}x}\xspace}
\newcommand{\RpPb}         {\ensuremath{R_{\rm pPb}}\xspace}

\newcommand{\nineH}        {$\sqrt{s}~=~0.9$~Te\kern-.1emV\xspace}
\newcommand{\seven}        {$\sqrt{s}~=~7$~Te\kern-.1emV\xspace}
\newcommand{\twoH}         {$\sqrt{s}~=~0.2$~Te\kern-.1emV\xspace}
\newcommand{\twosevensix}  {$\sqrt{s}~=~2.76$~Te\kern-.1emV\xspace}
\newcommand{\five}         {$\sqrt{s}~=~5.02$~Te\kern-.1emV\xspace}
\newcommand{\twosevensixnn}{$\sqrt{s_{\mathrm{NN}}}~=~2.76$~Te\kern-.1emV\xspace}
\newcommand{\fivenn}       {$\sqrt{s_{\mathrm{NN}}}~=~5.02$~Te\kern-.1emV\xspace}
\newcommand{\LT}           {L{\'e}vy-Tsallis\xspace}
\newcommand{\GeVc}         {Ge\kern-.1emV/$c$\xspace}
\newcommand{\MeVc}         {Me\kern-.1emV/$c$\xspace}
\newcommand{\TeV}          {Te\kern-.1emV\xspace}
\newcommand{\GeV}          {Ge\kern-.1emV\xspace}
\newcommand{\MeV}          {Me\kern-.1emV\xspace}
\newcommand{\GeVmass}      {Ge\kern-.2emV/$c^2$\xspace}
\newcommand{\MeVmass}      {Me\kern-.2emV/$c^2$\xspace}
\newcommand{\lumi}         {\ensuremath{\mathcal{L}}\xspace}

\newcommand{\ITS}          {\rm{ITS}\xspace}
\newcommand{\TOF}          {\rm{TOF}\xspace}
\newcommand{\ZDC}          {\rm{ZDC}\xspace}
\newcommand{\ZDCs}         {\rm{ZDCs}\xspace}
\newcommand{\ZNA}          {\rm{ZNA}\xspace}
\newcommand{\ZNC}          {\rm{ZNC}\xspace}
\newcommand{\SPD}          {\rm{SPD}\xspace}
\newcommand{\SDD}          {\rm{SDD}\xspace}
\newcommand{\SSD}          {\rm{SSD}\xspace}
\newcommand{\TPC}          {\rm{TPC}\xspace}
\newcommand{\TRD}          {\rm{TRD}\xspace}
\newcommand{\VZERO}        {\rm{V0}\xspace}
\newcommand{\VZEROA}       {\rm{V0A}\xspace}
\newcommand{\VZEROC}       {\rm{V0C}\xspace}
\newcommand{\Vdecay} 	   {\ensuremath{V^{0}}\xspace}

\newcommand{\hyp}         {${}^{3}_{\Lambda}\mathrm{H}\xspace}
\newcommand{\ee}           {\ensuremath{e^{+}e^{-}}} 
\newcommand{\pip}          {\ensuremath{\pi^{+}}\xspace}
\newcommand{\pim}          {\ensuremath{\pi^{-}}\xspace}
\newcommand{\kap}          {\ensuremath{\rm{K}^{+}}\xspace}
\newcommand{\kam}          {\ensuremath{\rm{K}^{-}}\xspace}
\newcommand{\pbar}         {\ensuremath{\rm\overline{p}}\xspace}
\newcommand{\kzero}        {\ensuremath{{\rm K}^{0}_{\rm{S}}}\xspace}
\newcommand{\lmb}          {\ensuremath{\Lambda}\xspace}
\newcommand{\almb}         {\ensuremath{\overline{\Lambda}}\xspace}
\newcommand{\Om}           {\ensuremath{\Omega^-}\xspace}
\newcommand{\Mo}           {\ensuremath{\overline{\Omega}^+}\xspace}
\newcommand{\X}            {\ensuremath{\Xi^-}\xspace}
\newcommand{\Ix}           {\ensuremath{\overline{\Xi}^+}\xspace}
\newcommand{\Xis}          {\ensuremath{\Xi^{\pm}}\xspace}
\newcommand{\Oms}          {\ensuremath{\Omega^{\pm}}\xspace}
\newcommand{\degree}       {\ensuremath{^{\rm o}}\xspace}

\newcommand{\detailtexcount}[1]{%
  \immediate\write18{texcount -merge -sum -q #1.tex output.bbl > #1.wcdetail }%
  \verbatiminput{#1.wcdetail}%
  \newpage
}

\newcommand{\quickwordcount}[1]{%
  \immediate\write18{texcount -1 -sum -merge -q #1.tex output.bbl > #1-words.sum }%
  \input{#1-words.sum} words%
}

\newcommand{\quickcharcount}[1]{%
  \immediate\write18{texcount -1 -sum -merge -char -q #1.tex output.bbl > #1-chars.sum }%
  \input{#1-chars.sum} characters (not including spaces)%
}

\begin{titlepage}
\PHyear{2022}           
\PHnumber{264}        
\PHdate{21 November}  

\title{Enhanced deuteron coalescence probability in jets}
\ShortTitle{Enhanced deuteron coalescence probability in jets}   

\Collaboration{ALICE Collaboration\thanks{See Appendix~\ref{app:collab} for the list of collaboration members}}
\ShortAuthor{ALICE Collaboration} 

\begin{abstract}
The transverse-momentum ($\pt$) spectra and coalescence parameters $B_2$ of (anti)deuterons are measured in pp collisions at $\sqrt{s} = $13 TeV for the first time in and out of jets. In this measurement, the direction of the leading particle with the highest $\pt$ in the event ($p_{\rm T}^{\rm{ lead}} >$ 5 GeV/$c$) is used as an approximation for the jet axis. The event is consequently divided into three azimuthal regions and the jet signal is obtained as the difference between the Toward region, that contains jet fragmentation products in addition to the underlying event (UE), and the Transverse region, which is dominated by the UE. The coalescence parameter in the jet is found to be approximately a factor of 10 larger than that in the underlying event. This experimental observation is consistent with the coalescence picture and can be attributed to the smaller average phase-space distance between nucleons in the jet cone as compared to the underlying event.
The results presented in this Letter are compared to predictions from a simple nucleon coalescence model, where the phase space distributions of nucleons are generated using PYTHIA 8 with the Monash 2013 tuning, and to predictions from a deuteron production model based on ordinary nuclear reactions with parametrized energy-dependent cross sections tuned on data. The latter model is implemented in PYTHIA 8.3. Both models reproduce the observed large difference between in-jet and out-of-jet coalescence parameters, although the almost flat trend of the $B^{\rm Jet}_2$ is not reproduced by the models, which instead give a decreasing trend. 
\end{abstract}
\end{titlepage}

\setcounter{page}{2}  

The production mechanisms of light (anti)nuclei in high-energy hadronic collisions is still unclear and currently under intense debate in the scientific community, despite the plethora of results published in recent years at different collision energies, ranging from the AGS~\cite{ Bennett:1998be, Ahle:1999in, Armstrong:2000gz, Armstrong:2000gd}, to the SPS~\cite{ Ambrosini:1997bf}, RHIC~\cite{Adler:2001prl,Adler:2004uy, Arsene:2010px,Agakishiev:2011ib,Adamczyk:2016gfs,Adam:2019wnb}, and the LHC~\cite{Adam:2015pna,nuclei_pp_PbPb,deuteron_pp_7TeV,deuteron_PbPb_276TeV,nuclei_pp,4He_PbPb,deuteron_pPbALICE,3He_pPb,deuteron_pp_13TeV,nuclei_pp_13TeV,deuteron_jets,nuclei_pp_5TeV,hypertriton_pPb_ALICE,deuteron_flow_276TeV,deuteron_flow_5TeV,helium3_flow_5TeV}. 
The experimental data are typically described using two different phenomenological approaches: the baryon coalescence approach and the statistical hadronization model (SHM). 
In the coalescence model~\cite{Coalescence1,Coalescence2,Coalescence3,Coalescence4,Coalescence5,Coalescence6,Coalescence7}, multi-baryon states are assumed to be formed by coalescence of baryons that are close in phase space at kinetic freeze-out (occurring when the elastic scatterings among the particles produced in the collision cease). In the simple coalescence approach~\cite{Coalescence2}, only momentum correlations are considered, while in state-of-the-art implementations~\cite{Coalescence3}, the quantum-mechanical properties of baryons and bound states are taken into account and the coalescence probability is calculated from the overlap between the phase-space distributions of individual (point-like) baryons and the Wigner density of the final-state cluster. The phase-space distributions of nucleons at freeze-out are generated using general-purpose Monte Carlo (MC) event generators or relativistic hydrodynamical simulations~\cite{Coalescence4, PhysRevC.99.044907}.
In the SHM~\cite{SHM1,SHM2,SHM3,SHM4,SHM5,SHM6,SHM7,SHM8}, light (anti)nuclei, as well as other hadron species, are assumed to be emitted by a source in local thermal and hadrochemical equilibrium with their abundances being fixed at chemical freeze-out (occurring when the inelastic scatterings stop). 
The grand-canonical approach provides an excellent description of the measured hadron yields in central nucleus--nucleus collisions, in which a large number of particles is produced, ranging from a few hundred to 2000 charged particles per unit of rapidity with increasing collision energy~\cite{SHM5}. 
For small systems, such as those produced in proton-proton (pp) and proton--nucleus collisions, the production of light nuclei can be described using a different implementation of this model based on the canonical ensemble, where exact conservation of quantum numbers is required~\cite{SHM8}. Significant deviations in this case are observed between the experimental data and the predictions from the canonical SHM~\cite{hypertriton_pPb_ALICE}. 
The study of the production mechanisms of multi-baryon bound states in high-energy hadronic collisions is one of the goals of the experimental program of ALICE. 

Measurements of light (anti)nuclei produced at accelerators are also important for indirect dark matter (DM) searches as they provide input for the estimates of the background of antinuclei produced in space. 
Several experiments, such as AMS-02~\cite{Kounine:2012ega} and GAPS~\cite{Hailey:2009fpa}, are searching for antinuclei~\cite{Donato:2008yx,Korsmeier:2017xzj}, but, so far, only antiprotons have been detected in space~\cite{Aguilar:2016kjl}. 
The possible presence of antinuclei in our Galaxy could be explained either by reactions of cosmic rays (CRs) with the interstellar medium (ISM) or by decays/annihilations of dark matter candidates. Both CRs and the ISM mostly consist of hydrogen (90$\%$), helium (8$\%$), and only in small percentage of heavier nuclei~\cite{antideuteronInelCS}. Hence, most of the relevant collisions for the production of CR antinuclei in the Galaxy are pp, p–$^4$He and $^4$He–p. Antinuclei are produced if the center-of-mass energy per nucleon--nucleon collision (\snn) of such collisions is above a given energy threshold, which is 6 GeV for antideuterons and 8 GeV for anti-$^3$He, for instance.
Hence, one of the goals of the existing experimental programs at different accelerator facilities is to pin down the microscopic production process of antinuclei in hadronic collisions or DM decay, in order to be able to correctly interpret the future measurements in space. 

Further insight into the (anti)nuclei production mechanisms can be obtained by measuring their production in and out of jets in pp collisions. Hadrons in the jet cone are closer in phase space compared with hadrons out of the jet. In a jet fragmentation, the produced particles are strongly correlated in phase space, i.e., particles which are close in space have also similar momenta~\cite{Field:1977fa, Andersson:1983ia}. 
In the coalescence picture, this condition should result in a larger coalescence probability in jets compared with that out of jets, since the proximity in both space and momentum is required for coalescence. Hence, the probability of coalescence in jets is increased with respect to that in the UE, where space and momentum are not correlated, i.e., close-by particles can have very different momenta and vice-versa.
The coalescence probability can be quantified by the coalescence parameter $B_A$, defined, in the case of deuterons, as the ratio between the invariant yield of the deuterons and the square of the invariant yield of protons: 

\begin{equation}
B_{2} = { \biggl( \dfrac{3}{2 \pi p^{\mathrm d}_{\mathrm T}} \dfrac{ \mathrm{d}^2N_{\mathrm d}}{\mathrm{d}y\mathrm{d} p_{\mathrm T}^{\mathrm d}}  \biggr)}  \bigg/{  \biggl( \dfrac{3}{2 \pi p^{\rm p}_{\mathrm T}} \dfrac{\mathrm{d}^2N_{\mathrm p} }{\mathrm{d}y\mathrm{d}p_{\mathrm T}^{\mathrm p}} \biggr)^2},
\label{eq:BA}
\end{equation}

\noindent where the labels d and p indicate the deuteron and the proton, respectively, and \mbox{$p_{\mathrm T}^{\rm p}$ = $p_{\mathrm T}^{\mathrm d}$/$2$}, assuming that protons and neutrons have the same production spectra, since they belong to the same isospin doublet.

Deuteron production in jets has already been measured in pp collisions at $\sqrt{s} = $13 TeV by the ALICE Collaboration using the two-particle correlation method~\cite{deuteron_jets}. The $p_{\mathrm{T}}$ spectrum of (anti)deuterons in the jet is found to be consistent with predictions based on the PYTHIA 8 event generator~\cite{Sjostrand:2007gs} coupled to a simple coalescence afterburner~\cite{coalescence_afterburner}. In the latter, deuteron formation is assumed to happen if a proton and a neutron have a momentum difference $\Delta p<p_{0}$, with $p_{0}$ = 110 MeV$/c$, thus ignoring the space coordinates in the calculation of the coalescence probability. It was also concluded that the yield of deuterons in jets, in events with a leading particle with \pt$>$5 GeV/$c$ at midrapidity, is about 10$\%$ of that in the underlying event (UE). 

This study is expanded in this Letter and complemented by the first measurement of (anti)deuteron coalescence parameters in and out of jets. The direction of a leading particle with high transverse-momentum ($p_{\rm T}^{\rm{ lead}} >$ 5 GeV/$c$) at midrapidity is used as an approximation for the jet axis. The \pt spectra and coalescence parameters of (anti)deuterons are measured in three different azimuthal regions defined based on $\Delta\varphi = |\varphi_{\rm d}  - \varphi_{\rm lead} |$, where $\varphi_{\rm d}$ is the azimuthal angle of the deuterons and $\varphi_{\rm lead}$ is that of the leading particle. The three $120^{\circ}$-wide azimuthal regions and the corresponding $\Delta\varphi$ intervals are the same as those defined in Refs.~\cite{UEpp, ALICE:2022fnb}: the one around the leading particle (Toward, $|\Delta\varphi |<60^{\circ}$), the one back-to-back to it (Away, $|\Delta\varphi |>120^{\circ}$), and the one transverse to both of them (Transverse, $60^{\circ}<|\Delta\varphi |<120^{\circ}$). The Toward and Away regions contain contributions from the leading and recoil jets in addition to the underlying event, while the Transverse region is dominated by the underlying event.

A detailed description of the ALICE apparatus and its performance can be found in Refs.~\cite{ALICEexperiment} and~\cite{ALICEperformance}. 
Trajectories of charged particles are reconstructed in the ALICE central barrel with the Inner Tracking System (ITS)~\cite{ALICEexperiment} and the Time Projection Chamber (TPC)~\cite{TPC}, which cover the full azimuthal angle and the pseudorapidity interval $|\eta|<$ 0.9. These are located within a large solenoidal magnet, providing a highly homogeneous magnetic field of 0.5 T parallel to the beam line. 
The TPC provides up to 159 spatial points per track for charged-particle reconstruction and particle identification (PID) through the measurement of the specific ionization energy loss d$E$/d$x$ in the gas volume. The PID is complemented by the Time-Of-Flight (TOF) system~\cite{TOF} located at a radial distance of 3.7 m from the nominal interaction point. It measures the arrival time of particles relative to the event collision time provided by the TOF detector itself or by the T0 detectors, two arrays of Cherenkov counters located at forward and backward rapidities (4.6 $<\eta<$ 4.9 and --3.3 $<\eta<$ --3.0)~\cite{CollisionTime, ALICEperformance}. 

The data used for this analysis were collected in 2016, 2017, and 2018 during the LHC pp runs at \mbox{$\sqrt{s}$ = 13 TeV}. A minimum bias (MB) event trigger was used, which requires coincident signals in the V0 detectors (two plastic scintillator arrays located at forward and backward rapidities, $2.8 < \eta < 5.1$ and $-3.7 < \eta < -1.7$~\cite{T0Cortese:2004aa}) synchronous with the bunch crossing time defined by the LHC clock. 
Events with multiple collision vertices, reconstructed from track segments in the two innermost layers of the ITS, are tagged as pile-up and removed from the analysis~\cite{ALICEperformance}. A total of approximately 1.7 billion MB pp events are considered for further analysis, corresponding to an integrated luminosity of about 30 nb$^{-1}$. Events with at least one charged primary particle with \pt $>$ 5 GeV/$c$ are selected, which correspond to approximately 1$\%$ of the MB pp collisions.

Deuteron candidates are selected from charged-particle tracks reconstructed in the ITS and TPC in the pseudorapidity interval $|\eta|<0.8$. Basic track quality criteria are applied, e.g., selecting on the number of space points associated to the track in the TPC and ITS and on the $\chi^2$ of the track fit, as done in previous light (anti)nucleus analyses, such as in Ref.~\cite{deuteron_pp_13TeV}. 
The track selection criteria used for the leading particle are the same as in Ref.~\cite{UEpp}. 

(Anti)deuterons are identified in the rapidity interval $|y|<$ 0.5 in the range $0.6 <$ \pt  $< 1.2\ \mathrm{GeV}/\textit{c}$ by requiring that their energy loss per unit of track length measured by the TPC is within 3$\sigma_{\mathrm dE/\mathrm dx}$ from the expected average for (anti)deuterons, where $\sigma_{\mathrm dE/\mathrm dx}$ is the \dEdx resolution. For $1.2 <$ \pt $< 3.0\ \mathrm{GeV}/\textit{c}$, the particle identification is complemented by the TOF and the (anti)deuteron signal is extracted from a fit to the n$\sigma_{\rm{TOF}}$ = $(\Delta t - \Delta t_{\mathrm{d}}) / \sigma_{\mathrm{TOF}}$ distribution, where $\Delta t$ is the measured time-of-flight, $\Delta t_{\mathrm{d}}$ its expected value for deuterons and $\sigma_{\mathrm{TOF}}$ the resolution on the time-of-flight measurement. The fit function consists of a Gaussian with an exponential tail on the right side for the signal and the sum of two exponential functions for the background~\cite{nuclei_pp_PbPb}. The signal is extracted by integrating the signal function in the asymmetric interval [$\mu_{0}-3\sigma_{\mathrm{TOF}}, \mu_{0}+3.5\sigma_{\mathrm{TOF}}$], where $\mu_{0}$ is the mean of the Gaussian function. 

The raw (anti)deuteron \pt spectra are corrected for the reconstruction efficiency and, only for deuterons, for the fraction of secondary deuterons produced by spallation in interactions between primary particles and the detector material. This correction is based on fits to the distance of closest approach (DCA) of the deuteron track to the primary vertex using template distributions extracted from MC simulations, as described in Refs.~\cite{deuteron_pp_13TeV,nuclei_pp_13TeV,nuclei_pp_5TeV}. Both corrections are calculated using MC simulations in which (anti)nuclei are embedded into pp collision events generated using PYTHIA 8.1 with the Monash 2013 tune~\cite{PYTHIA8Monasch2013}. (Anti)nuclei are generated with uniform \pt and rapidity distributions within \mbox{0 $<$ \pt $<$ 10 $\mathrm{GeV}/\textit{c}$} and $-1 < y < 1$. The passage of particles through the ALICE detector is simulated using GEANT4~\cite{GEANT4}. The reconstruction efficiency is calculated as the ratio of the number of reconstructed and generated (anti)deuterons in the simulation, after the proper re-weighting of the generated \pt distribution according to the L\'{e}vy-Tsallis function. The same track selection and PID criteria as those used in the data are applied. The fraction of secondary deuterons is approximately 0.5 in the lowest \pt interval ($0.6 < p_{\rm T} < 0.8$ GeV/$c$) and decreases exponentially with increasing \pt, becoming negligible for $p_{\rm T} > 1.6$ GeV/$c$. The fraction of secondary deuterons is weakly dependent on the azimuthal region. 

The dominant contributions to the systematic uncertainties are related to the track selection, the particle identification, the limited knowledge of the ALICE material budget, the uncertainties on the hadronic interaction cross section of (anti)deuterons, and the procedure used to estimate the fraction of secondary nuclei.
The uncertainty related to the track selection is estimated by repeating the analysis using different track selection criteria and considering the RMS of the distribution of the results in each \pt interval. This contribution is found to range between 8 (1)$\%$ and 6 (5)$\%$ for deuterons (antideuterons), depending on \pt. Such uncertainty is larger for deuterons at low \pt, due to the DCA cut variations which are sensitive to the secondary nuclei. The PID uncertainty is estimated by varying the (anti)deuteron selection criteria in the TPC and TOF, the fit functions, and the signal extraction strategy (bin counting or integral of the fit function). The PID systematic uncertainty rises with \pt, from 4$\%$ to 8$\%$.
The systematic uncertainty due to the estimate of the primary fraction correction is obtained by varying the conditions (selection criteria, histogram bin width, fit range) of the fits to the DCA distributions used to estimate the correction factors. This contribution depends on the azimuthal region, and it ranges between 14$\%$ (10$\%$ and 7$\%$) and few percent for the Away region (Toward and Transverse, respectively).
The contribution from ITS--TPC and TPC--TOF matching efficiencies are estimated from charged particle tracks by comparing the probabilities of prolonging a track from the TPC to the ITS (TOF) in the data and MC~\cite{deuteron_pp_13TeV}. The uncertainty is approximately 2$\%$ for the ITS--TPC matching efficiency and 5$\%$ for the TPC--TOF one, weakly dependent on \pt.
The uncertainty on the material budget (1$\%$) is taken from Ref.~\cite{deuteron_pp_13TeV}. The uncertainty on the hadronic interaction cross section of (anti)deuterons is estimated by varying the default inelastic cross section of (anti)deuterons in GEANT4 by the uncertainties of the measurements~\cite{absorption1,absorption2,antideuteronAbs_13, antideuteronAbs_25, antideuteronInelCS} and is $\sim$1$\%$ (3$\%$) for deuterons (antideuterons). The total uncertainty is obtained as the quadratic sum of each individual contribution and is found to be about 10$\%$.

\begin{figure}[!hbt]
\centering
	\includegraphics[width=0.48\textwidth]{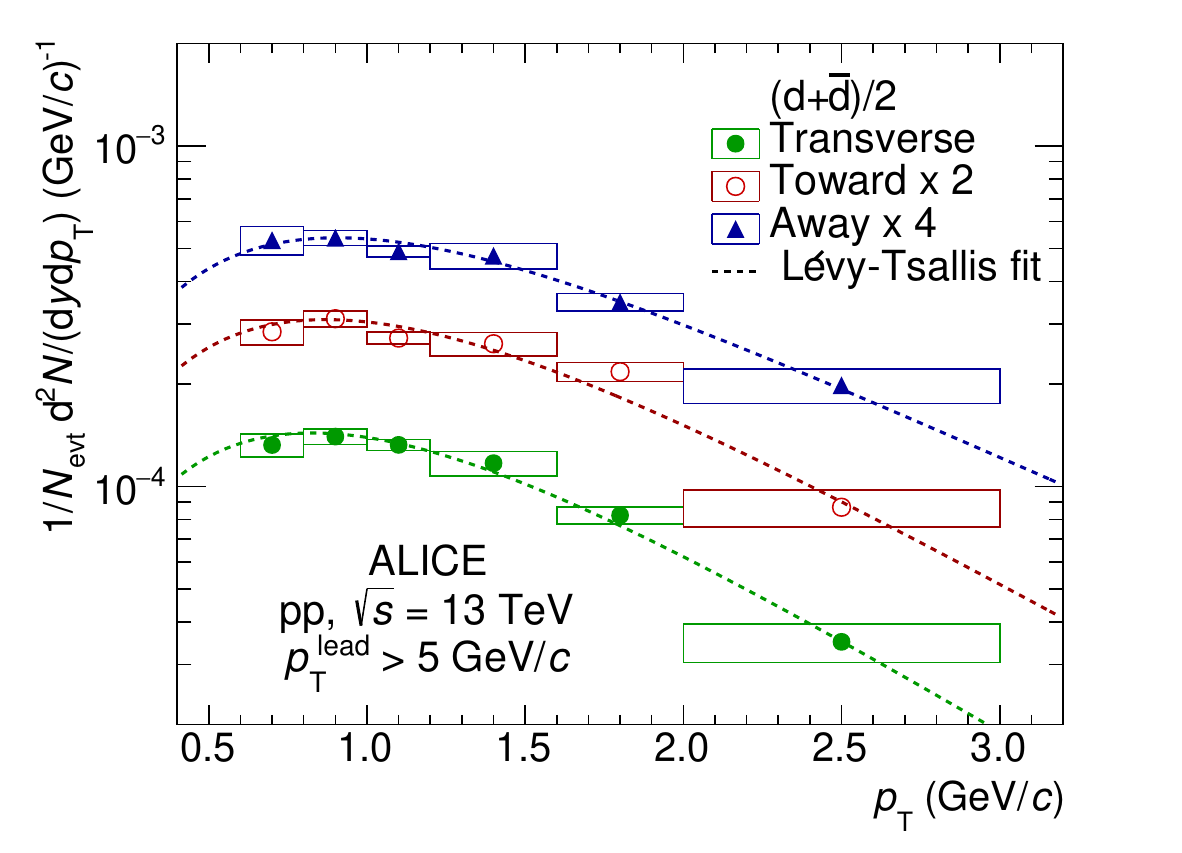}	
	\includegraphics[width=0.48\textwidth]{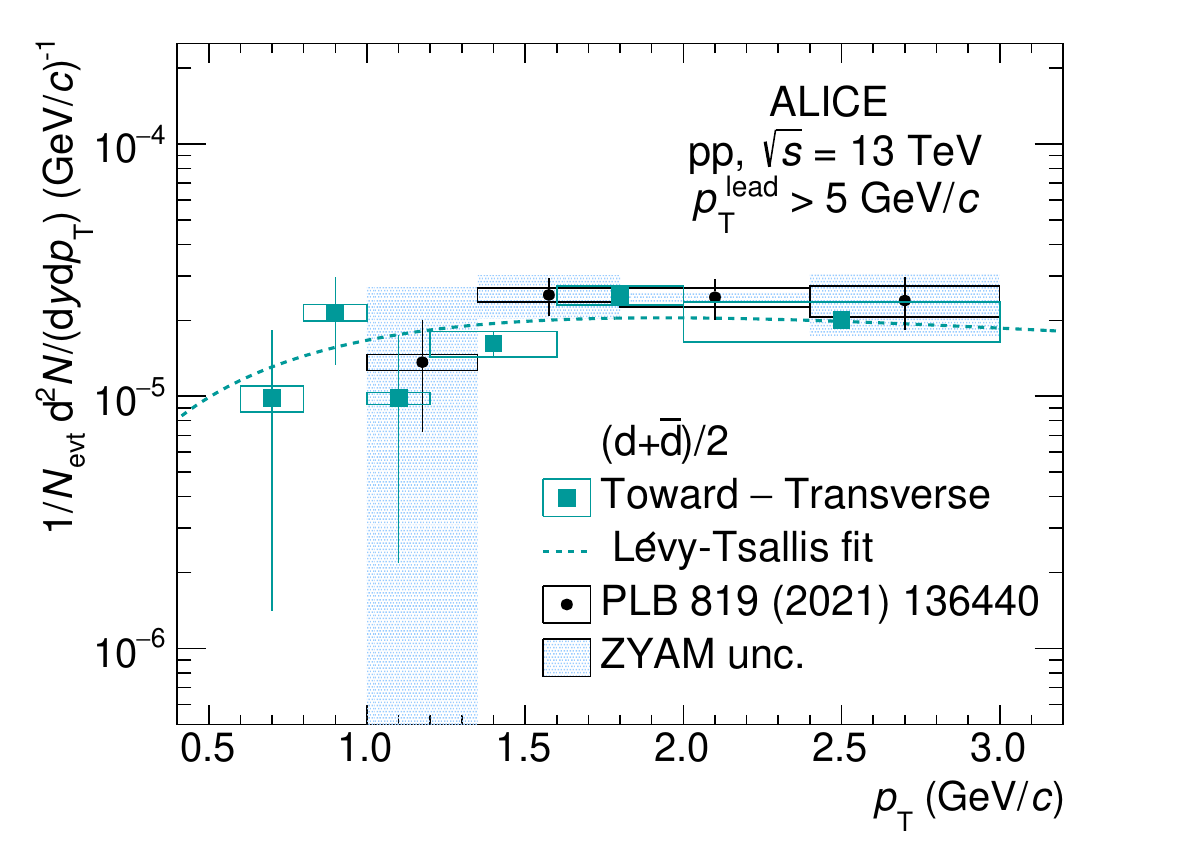}
\caption{Average of deuteron and antideuteron \pt-differential yields in the three azimuthal regions (on the left) and in jets (on the right), measured in pp collisions at $\sqrt{s} = $ 13 TeV. 
Statistical and systematic uncertainties are represented by vertical bars and boxes, respectively. On the right, shaded (blue) boxes show the uncertainty on the results from Ref.~\cite{deuteron_jets} related to the subtraction of the uncorrelated background using the ZYAM method~\cite{PHENIX:2005zfm}. Individual L{\'e}vy-Tsallis fits are also shown. }
\label{fig:pt_spectra}
\end{figure}

The average of the deuteron and antideuteron \pt distributions measured in the three azimuthal regions discussed above, are shown in Fig.~\ref{fig:pt_spectra} left. Individual L{\'e}vy-Tsallis fits~\cite{Tsallis:1987eu} are also shown. The \pt spectrum of (anti)deuterons in jets is obtained by subtracting the spectrum in the Transverse region, dominated by the underlying event, from that in the Toward region, which contains contributions from both jet fragmentation and the underlying event. In the subtraction, the systematic uncertainties are treated as independent, to be conservative. 
The resulting spectrum, with a much harder shape compared to the UE spectrum, is shown in Fig.~\ref{fig:pt_spectra} right, with squared markers, and compared to the in-jet deuteron spectrum measured with the two-particle correlation method taken from Ref.~\cite{deuteron_jets}. Excellent agreement is found between the deuteron \pt spectra in jets obtained with the two methods, proving the validity of the subtraction method to obtain the deuteron spectrum in jets. These results confirm that the integrated yield of deuterons in jets, in events with \mbox{$p_{\rm T}^{\rm{ lead}} >$ 5 GeV/$c$} at midrapidity, is only roughly 10$\%$ of that in the UE, as already pointed out in Ref.~\cite{deuteron_jets}. 

\begin{figure}[!hbt]
\centering
	\includegraphics[width = 0.48\textwidth]{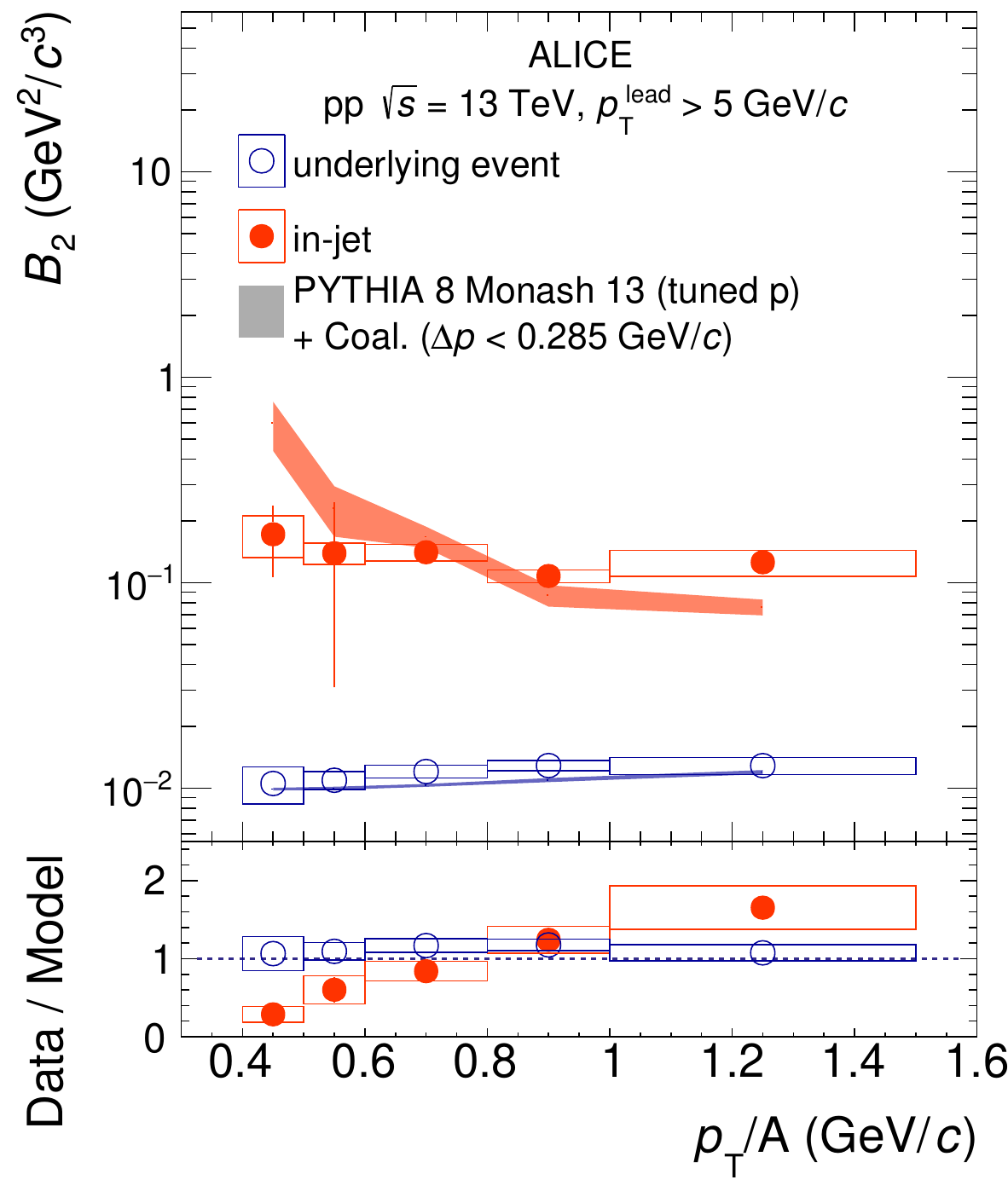}
	\includegraphics[width = 0.48\textwidth]{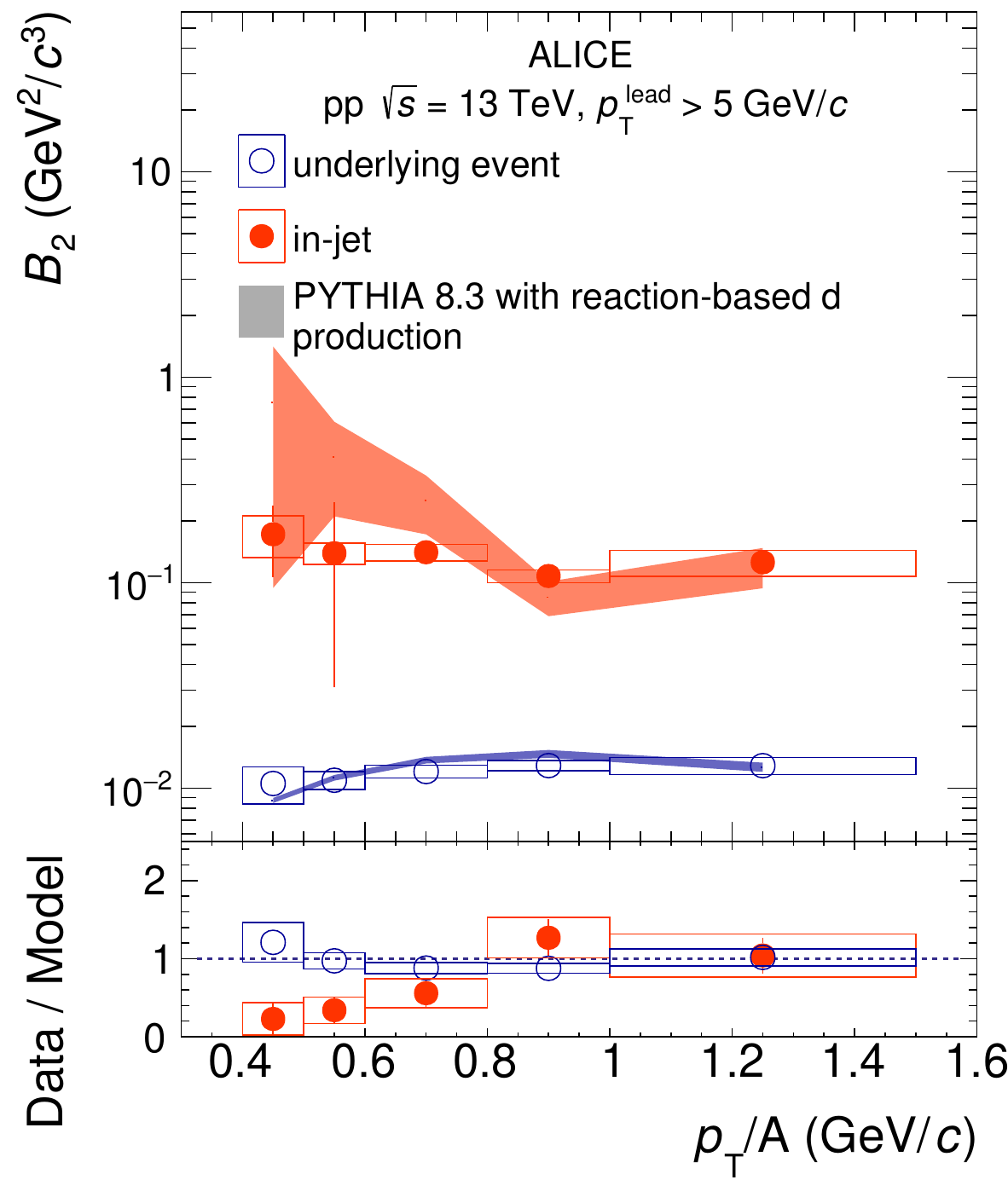}
\caption{Coalescence parameter in jets, $B^{\rm Jet}_{2}$, and in the underlying event, $B^{\rm UE}_{2}$, as a function of $p_{\rm T}/A$ in comparison with the predictions from PYTHIA 8 with the Monash 2013 tune with a coalescence afterburner (left panel) and the reaction-based PYTHIA 8.3 model (right panel).}
\label{fig:coalescenceB2}
\end{figure}

The coalescence parameters in the jet ($B^{\rm Jet}_2$) and underlying event ($B^{\rm UE}_2$) are calculated using Eq.~\ref{eq:BA}. 
The spectra for (anti)protons are derived from the $R_{\rm T}$-differential \pt spectra published in Ref.~\cite{ALICE:2023yuk}, integrating over the full $R_{\rm T}$, which is the underlying event activity classifier defined as $R_{\rm T}=N^{\rm ch}_{\rm T}/ \langle N^{\rm ch}_{\rm T} \rangle$ ($N^{\rm ch}_{\rm T}$ being the charged-particle multiplicity measured in the Transverse region and $\langle N^{\rm ch}_{\rm T} \rangle$ its average)~\cite{UEpp}.
The $p_{\rm T}$ spectrum of (anti)protons in jets is obtained by subtracting the efficiency-corrected \pt spectra in the Transverse region from those in the Toward region taken from Ref.~\cite{ALICE:2023yuk}, as done for deuterons. The coalescence parameters $B^{\rm Jet}_{2}$ and $B^{\rm UE}_2$ are shown in Fig.~\ref{fig:coalescenceB2} as a function of the transverse momentum per nucleon, $p_{\rm T}/A$. The coalescence parameter $B^{\rm Jet}_{2}$ is found to be about a factor of 10 larger than that in the underlying event, with a significance of about 21$\sigma$. Hence, despite the limited contribution of nuclei arising from jets to their total production yield, the probability that two nucleons in the jet cone coalesce into a deuteron is enhanced with respect to the corresponding coalescence probability in the underlying event. This experimental observation can be interpreted, within the coalescence model, as due to the reduced distance in phase space between nucleons in the jets compared to those in the underlying event.

These results are compared to predictions from simple coalescence and reaction-based deuteron production models. In the former approach, the phase-space distributions of nucleons are generated with PYTHIA 8 with the Monash 2013 tune~\cite{PYTHIA8Monasch2013}, and their \pt spectra are re-weighted to match the proton \pt spectra measured by ALICE in the three different azimuthal regions~\cite{ALICE:2023yuk}. Spatial correlations are ignored and deuteron formation is assumed to happen if a proton and a neutron have a momentum difference below a given coalescence momentum $\Delta p<p_{0}$ in the deuteron rest frame. The best estimate of the coalescence momentum, for this model, is $p_0=285 \pm 1$ MeV/$c$, which is consistent with the value of Ref.~\cite{PhysRevC.99.014901}. A detailed description of these coalescence calculations can be found in the Supplemental Material~\cite{Additional_Materal}. 

In the reaction-based model~\cite{Bierlich:2022pfr}, (anti)deuterons are generated by ordinary nuclear reactions between nucleons produced in the collision with parameterized energy-dependent cross sections tuned on available experimental data~\cite{Dal:2015sha}. 
The following nuclear reactions are considered: p + n$ \rightarrow \gamma$ + d, p + n$\rightarrow \pi^0 $ + d, p + n $\rightarrow \pi^0$ + $\pi^0$ + d, p + n $\rightarrow \pi^+$ + $\pi^-$ + d, p + p $\rightarrow \pi^+$ + d, p + p $\rightarrow \pi^+$ + $\pi^0$ + d, n + n $\rightarrow \pi^-$ + d, n + n $\rightarrow \pi^-$ + $\pi^0$ + d. 
Such a model is implemented in the MC event generator PYTHIA 8.3.
Both simple coalescence and the reaction-based model are based on the assumption that coalescence can happen only if protons and neutrons are close in phase space. However, there are two main differences between the models. The first concerns the $p_0$ cut off which is introduced as a step function in the coalescence model, while in PYTHIA 8.3 the deuteron formation happens according to the differential cross section of the reactions listed above. The second difference is related to the kinematics of the process: in the coalescence model the deuteron four-momentum is given by the sum of the proton and neutron four-momenta, while in the reaction-based model, part of the initial four-momentum is carried away by the pion(s) or photon according to energy and momentum conservation rules. The uncertainties of the coalescence model are discussed in the Supplemental Material~\cite{Additional_Materal}, while for PYTHIA 8.3 only statistical uncertainties are shown. 

As shown in Fig.~\ref{fig:coalescenceB2}, both models provide good qualitative descriptions of the data. In particular, they are capable of reproducing the observed large difference between $B^{\rm Jet}_2$ and $B^{\rm UE}_2$. While the \pt/$A$ dependence of $B^{\rm UE}_2$ is well described by both approaches within the uncertainties, the observed nearly flat trend of $B^{\rm Jet}_2$ is not reproduced by the models, which instead give a decreasing trend with increasing \pt/$A$ and overestimate the $B^{\rm Jet}_2$ for low \pt/$A$. 

In this Letter the first measurement of the (anti)deuteron coalescence parameters in and out of jets in pp collisions at \mbox{$\sqrt{s}$ = 13 TeV} is presented. The results of this work indicate for the first time an enhanced deuteron coalescence probability in jets compared to the underlying event. This enhancement, of a factor \mbox{$\sim$ 10}, is measured with a good precision and a significance of about 21$\sigma$. This experimental observation is a parameter-free and absolute prediction of coalescence, which decisively proves the formation of bound states by coalescence when nucleons have a smaller average phase-space distance, as it is the case in the jet cone.
This study will be extended to heavier (anti)nuclei, such as (anti)$^{3}$He, and to p--Pb collisions, with the LHC Run 3 data. 
Further investigations of the coalescence parameters $B_A$ in the jet and in the underlying event, in small collision systems, which reproduce the CR interactions originating nuclei in our Galaxy, will provide additional insight into the production mechanisms and will contribute to further constrain the coalescence models. These studies will be crucial to correctly interpret any future measurement of antinuclei in satellite and balloon-borne experiments in space. 
       


\newenvironment{acknowledgement}{\relax}{\relax}
\begin{acknowledgement}
\section*{Acknowledgements}

The ALICE Collaboration would like to thank all its engineers and technicians for their invaluable contributions to the construction of the experiment and the CERN accelerator teams for the outstanding performance of the LHC complex.
The ALICE Collaboration gratefully acknowledges the resources and support provided by all Grid centres and the Worldwide LHC Computing Grid (WLCG) collaboration.
The ALICE Collaboration acknowledges the following funding agencies for their support in building and running the ALICE detector:
A. I. Alikhanyan National Science Laboratory (Yerevan Physics Institute) Foundation (ANSL), State Committee of Science and World Federation of Scientists (WFS), Armenia;
Austrian Academy of Sciences, Austrian Science Fund (FWF): [M 2467-N36] and Nationalstiftung f\"{u}r Forschung, Technologie und Entwicklung, Austria;
Ministry of Communications and High Technologies, National Nuclear Research Center, Azerbaijan;
Conselho Nacional de Desenvolvimento Cient\'{\i}fico e Tecnol\'{o}gico (CNPq), Financiadora de Estudos e Projetos (Finep), Funda\c{c}\~{a}o de Amparo \`{a} Pesquisa do Estado de S\~{a}o Paulo (FAPESP) and Universidade Federal do Rio Grande do Sul (UFRGS), Brazil;
Bulgarian Ministry of Education and Science, within the National Roadmap for Research Infrastructures 2020-2027 (object CERN), Bulgaria;
Ministry of Education of China (MOEC) , Ministry of Science \& Technology of China (MSTC) and National Natural Science Foundation of China (NSFC), China;
Ministry of Science and Education and Croatian Science Foundation, Croatia;
Centro de Aplicaciones Tecnol\'{o}gicas y Desarrollo Nuclear (CEADEN), Cubaenerg\'{\i}a, Cuba;
Ministry of Education, Youth and Sports of the Czech Republic, Czech Republic;
The Danish Council for Independent Research | Natural Sciences, the VILLUM FONDEN and Danish National Research Foundation (DNRF), Denmark;
Helsinki Institute of Physics (HIP), Finland;
Commissariat \`{a} l'Energie Atomique (CEA) and Institut National de Physique Nucl\'{e}aire et de Physique des Particules (IN2P3) and Centre National de la Recherche Scientifique (CNRS), France;
Bundesministerium f\"{u}r Bildung und Forschung (BMBF) and GSI Helmholtzzentrum f\"{u}r Schwerionenforschung GmbH, Germany;
General Secretariat for Research and Technology, Ministry of Education, Research and Religions, Greece;
National Research, Development and Innovation Office, Hungary;
Department of Atomic Energy Government of India (DAE), Department of Science and Technology, Government of India (DST), University Grants Commission, Government of India (UGC) and Council of Scientific and Industrial Research (CSIR), India;
National Research and Innovation Agency - BRIN, Indonesia;
Istituto Nazionale di Fisica Nucleare (INFN), Italy;
Japanese Ministry of Education, Culture, Sports, Science and Technology (MEXT) and Japan Society for the Promotion of Science (JSPS) KAKENHI, Japan;
Consejo Nacional de Ciencia (CONACYT) y Tecnolog\'{i}a, through Fondo de Cooperaci\'{o}n Internacional en Ciencia y Tecnolog\'{i}a (FONCICYT) and Direcci\'{o}n General de Asuntos del Personal Academico (DGAPA), Mexico;
Nederlandse Organisatie voor Wetenschappelijk Onderzoek (NWO), Netherlands;
The Research Council of Norway, Norway;
Commission on Science and Technology for Sustainable Development in the South (COMSATS), Pakistan;
Pontificia Universidad Cat\'{o}lica del Per\'{u}, Peru;
Ministry of Education and Science, National Science Centre and WUT ID-UB, Poland;
Korea Institute of Science and Technology Information and National Research Foundation of Korea (NRF), Republic of Korea;
Ministry of Education and Scientific Research, Institute of Atomic Physics, Ministry of Research and Innovation and Institute of Atomic Physics and University Politehnica of Bucharest, Romania;
Ministry of Education, Science, Research and Sport of the Slovak Republic, Slovakia;
National Research Foundation of South Africa, South Africa;
Swedish Research Council (VR) and Knut \& Alice Wallenberg Foundation (KAW), Sweden;
European Organization for Nuclear Research, Switzerland;
Suranaree University of Technology (SUT), National Science and Technology Development Agency (NSTDA), Thailand Science Research and Innovation (TSRI) and National Science, Research and Innovation Fund (NSRF), Thailand;
Turkish Energy, Nuclear and Mineral Research Agency (TENMAK), Turkey;
National Academy of  Sciences of Ukraine, Ukraine;
Science and Technology Facilities Council (STFC), United Kingdom;
National Science Foundation of the United States of America (NSF) and United States Department of Energy, Office of Nuclear Physics (DOE NP), United States of America.
In addition, individual groups or members have received support from:
Marie Sk\l{}odowska Curie, European Research Council, Strong 2020 - Horizon 2020 (grant nos. 950692, 824093, 896850), European Union;
Academy of Finland (Center of Excellence in Quark Matter) (grant nos. 346327, 346328), Finland;
Programa de Apoyos para la Superaci\'{o}n del Personal Acad\'{e}mico, UNAM, Mexico.

\end{acknowledgement}

\bibliographystyle{utphys}   
\bibliography{bibliography}

\newpage
\appendix

\section{The ALICE Collaboration}
\label{app:collab}
\begin{flushleft} 
\small

S.~Acharya\,\orcidlink{0000-0002-9213-5329}\,$^{\rm 125}$, 
D.~Adamov\'{a}\,\orcidlink{0000-0002-0504-7428}\,$^{\rm 86}$, 
A.~Adler$^{\rm 69}$, 
G.~Aglieri Rinella\,\orcidlink{0000-0002-9611-3696}\,$^{\rm 32}$, 
M.~Agnello\,\orcidlink{0000-0002-0760-5075}\,$^{\rm 29}$, 
N.~Agrawal\,\orcidlink{0000-0003-0348-9836}\,$^{\rm 50}$, 
Z.~Ahammed\,\orcidlink{0000-0001-5241-7412}\,$^{\rm 132}$, 
S.~Ahmad\,\orcidlink{0000-0003-0497-5705}\,$^{\rm 15}$, 
S.U.~Ahn\,\orcidlink{0000-0001-8847-489X}\,$^{\rm 70}$, 
I.~Ahuja\,\orcidlink{0000-0002-4417-1392}\,$^{\rm 37}$
% ...

\section*{Affiliation Notes}

$^{\rm I}$ Deceased\\
$^{\rm II}$ Also at: Max-Planck-Institut f\"{u}r Physik, Munich, Germany\\
$^{\rm III}$ Also at: Italian National Agency for New Technologies, Energy and Sustainable Economic Development (ENEA), Bologna, Italy\\
$^{\rm IV}$ Also at: Dipartimento DET del Politecnico di Torino, Turin, Italy\\
$^{\rm V}$ Also at: Department of Applied Physics, Aligarh Muslim University, Aligarh, India\\
$^{\rm VI}$ Also at: Institute of Theoretical Physics, University of Wroclaw, Poland\\
$^{\rm VII}$ Also at: An institution covered by a cooperation agreement with CERN\\

\section*{Collaboration Institutes}

$^{1}$ A.I. Alikhanyan National Science Laboratory (Yerevan Physics Institute) Foundation, Yerevan, Armenia\\
$^{2}$ AGH University of Science and Technology, Cracow, Poland\\
$^{3}$ Bogolyubov Institute for Theoretical Physics, National Academy of Sciences of Ukraine, Kiev, Ukraine\\
$^{4}$ Bose Institute, Department of Physics  and Centre for Astroparticle Physics and Space Science (CAPSS), Kolkata, India\\
$^{5}$ California Polytechnic State University, San Luis Obispo, California, United States\\
$^{6}$ Central China Normal University, Wuhan, China\\
$^{7}$ Centro de Aplicaciones Tecnol\'{o}gicas y Desarrollo Nuclear (CEADEN), Havana, Cuba\\
$^{8}$ Centro de Investigaci\'{o}n y de Estudios Avanzados (CINVESTAV), Mexico City and M\'{e}rida, Mexico\\
$^{9}$ Chicago State University, Chicago, Illinois, United States\\
$^{10}$ China Institute of Atomic Energy, Beijing, China\\
$^{11}$ Chungbuk National University, Cheongju, Republic of Korea\\
$^{12}$ Comenius University Bratislava, Faculty of Mathematics, Physics and Informatics, Bratislava, Slovak Republic\\
$^{13}$ COMSATS University Islamabad, Islamabad, Pakistan\\
$^{14}$ Creighton University, Omaha, Nebraska, United States\\
$^{15}$ Department of Physics, Aligarh Muslim University, Aligarh, India\\
$^{16}$ Department of Physics, Pusan National University, Pusan, Republic of Korea\\
$^{17}$ Department of Physics, Sejong University, Seoul, Republic of Korea\\
$^{18}$ Department of Physics, University of California, Berkeley, California, United States\\
$^{19}$ Department of Physics, University of Oslo, Oslo, Norway\\
$^{20}$ Department of Physics and Technology, University of Bergen, Bergen, Norway\\
$^{21}$ Dipartimento di Fisica, Universit\`{a} di Pavia, Pavia, Italy\\
$^{22}$ Dipartimento di Fisica dell'Universit\`{a} and Sezione INFN, Cagliari, Italy\\
$^{23}$ Dipartimento di Fisica dell'Universit\`{a} and Sezione INFN, Trieste, Italy\\
$^{24}$ Dipartimento di Fisica dell'Universit\`{a} and Sezione INFN, Turin, Italy\\
$^{25}$ Dipartimento di Fisica e Astronomia dell'Universit\`{a} and Sezione INFN, Bologna, Italy\\
$^{26}$ Dipartimento di Fisica e Astronomia dell'Universit\`{a} and Sezione INFN, Catania, Italy\\
$^{27}$ Dipartimento di Fisica e Astronomia dell'Universit\`{a} and Sezione INFN, Padova, Italy\\
$^{28}$ Dipartimento di Fisica `E.R.~Caianiello' dell'Universit\`{a} and Gruppo Collegato INFN, Salerno, Italy\\
$^{29}$ Dipartimento DISAT del Politecnico and Sezione INFN, Turin, Italy\\
$^{30}$ Dipartimento di Scienze MIFT, Universit\`{a} di Messina, Messina, Italy\\
$^{31}$ Dipartimento Interateneo di Fisica `M.~Merlin' and Sezione INFN, Bari, Italy\\
$^{32}$ European Organization for Nuclear Research (CERN), Geneva, Switzerland\\
$^{33}$ Faculty of Electrical Engineering, Mechanical Engineering and Naval Architecture, University of Split, Split, Croatia\\
$^{34}$ Faculty of Engineering and Science, Western Norway University of Applied Sciences, Bergen, Norway\\
$^{35}$ Faculty of Nuclear Sciences and Physical Engineering, Czech Technical University in Prague, Prague, Czech Republic\\
$^{36}$ Faculty of Physics, Sofia University, Sofia, Bulgaria\\
$^{37}$ Faculty of Science, P.J.~\v{S}af\'{a}rik University, Ko\v{s}ice, Slovak Republic\\
$^{38}$ Frankfurt Institute for Advanced Studies, Johann Wolfgang Goethe-Universit\"{a}t Frankfurt, Frankfurt, Germany\\
$^{39}$ Fudan University, Shanghai, China\\
$^{40}$ Gangneung-Wonju National University, Gangneung, Republic of Korea\\
$^{41}$ Gauhati University, Department of Physics, Guwahati, India\\
$^{42}$ Helmholtz-Institut f\"{u}r Strahlen- und Kernphysik, Rheinische Friedrich-Wilhelms-Universit\"{a}t Bonn, Bonn, Germany\\
$^{43}$ Helsinki Institute of Physics (HIP), Helsinki, Finland\\
$^{44}$ High Energy Physics Group,  Universidad Aut\'{o}noma de Puebla, Puebla, Mexico\\
$^{45}$ Horia Hulubei National Institute of Physics and Nuclear Engineering, Bucharest, Romania\\
$^{46}$ Indian Institute of Technology Bombay (IIT), Mumbai, India\\
$^{47}$ Indian Institute of Technology Indore, Indore, India\\
$^{48}$ INFN, Laboratori Nazionali di Frascati, Frascati, Italy\\
$^{49}$ INFN, Sezione di Bari, Bari, Italy\\
$^{50}$ INFN, Sezione di Bologna, Bologna, Italy\\
$^{51}$ INFN, Sezione di Cagliari, Cagliari, Italy\\
$^{52}$ INFN, Sezione di Catania, Catania, Italy\\
$^{53}$ INFN, Sezione di Padova, Padova, Italy\\
$^{54}$ INFN, Sezione di Pavia, Pavia, Italy\\
$^{55}$ INFN, Sezione di Torino, Turin, Italy\\
$^{56}$ INFN, Sezione di Trieste, Trieste, Italy\\
$^{57}$ Inha University, Incheon, Republic of Korea\\
$^{58}$ Institute for Gravitational and Subatomic Physics (GRASP), Utrecht University/Nikhef, Utrecht, Netherlands\\
$^{59}$ Institute of Experimental Physics, Slovak Academy of Sciences, Ko\v{s}ice, Slovak Republic\\
$^{60}$ Institute of Physics, Homi Bhabha National Institute, Bhubaneswar, India\\
$^{61}$ Institute of Physics of the Czech Academy of Sciences, Prague, Czech Republic\\
$^{62}$ Institute of Space Science (ISS), Bucharest, Romania\\
$^{63}$ Institut f\"{u}r Kernphysik, Johann Wolfgang Goethe-Universit\"{a}t Frankfurt, Frankfurt, Germany\\
$^{64}$ Instituto de Ciencias Nucleares, Universidad Nacional Aut\'{o}noma de M\'{e}xico, Mexico City, Mexico\\
$^{65}$ Instituto de F\'{i}sica, Universidade Federal do Rio Grande do Sul (UFRGS), Porto Alegre, Brazil\\
$^{66}$ Instituto de F\'{\i}sica, Universidad Nacional Aut\'{o}noma de M\'{e}xico, Mexico City, Mexico\\
$^{67}$ iThemba LABS, National Research Foundation, Somerset West, South Africa\\
$^{68}$ Jeonbuk National University, Jeonju, Republic of Korea\\
$^{69}$ Johann-Wolfgang-Goethe Universit\"{a}t Frankfurt Institut f\"{u}r Informatik, Fachbereich Informatik und Mathematik, Frankfurt, Germany\\
$^{70}$ Korea Institute of Science and Technology Information, Daejeon, Republic of Korea\\
$^{71}$ KTO Karatay University, Konya, Turkey\\
$^{72}$ Laboratoire de Physique des 2 Infinis, Ir\`{e}ne Joliot-Curie, Orsay, France\\
$^{73}$ Laboratoire de Physique Subatomique et de Cosmologie, Universit\'{e} Grenoble-Alpes, CNRS-IN2P3, Grenoble, France\\
$^{74}$ Lawrence Berkeley National Laboratory, Berkeley, California, United States\\
$^{75}$ Lund University Department of Physics, Division of Particle Physics, Lund, Sweden\\
$^{76}$ Nagasaki Institute of Applied Science, Nagasaki, Japan\\
$^{77}$ Nara Women{'}s University (NWU), Nara, Japan\\
$^{78}$ National and Kapodistrian University of Athens, School of Science, Department of Physics , Athens, Greece\\
$^{79}$ National Centre for Nuclear Research, Warsaw, Poland\\
$^{80}$ National Institute of Science Education and Research, Homi Bhabha National Institute, Jatni, India\\
$^{81}$ National Nuclear Research Center, Baku, Azerbaijan\\
$^{82}$ National Research and Innovation Agency - BRIN, Jakarta, Indonesia\\
$^{83}$ Niels Bohr Institute, University of Copenhagen, Copenhagen, Denmark\\
$^{84}$ Nikhef, National institute for subatomic physics, Amsterdam, Netherlands\\
$^{85}$ Nuclear Physics Group, STFC Daresbury Laboratory, Daresbury, United Kingdom\\
$^{86}$ Nuclear Physics Institute of the Czech Academy of Sciences, Husinec-\v{R}e\v{z}, Czech Republic\\
$^{87}$ Oak Ridge National Laboratory, Oak Ridge, Tennessee, United States\\
$^{88}$ Ohio State University, Columbus, Ohio, United States\\
$^{89}$ Physics department, Faculty of science, University of Zagreb, Zagreb, Croatia\\
$^{90}$ Physics Department, Panjab University, Chandigarh, India\\
$^{91}$ Physics Department, University of Jammu, Jammu, India\\
$^{92}$ Physics Program and International Institute for Sustainability with Knotted Chiral Meta Matter (SKCM2), Hiroshima University, Hiroshima, Japan\\
$^{93}$ Physikalisches Institut, Eberhard-Karls-Universit\"{a}t T\"{u}bingen, T\"{u}bingen, Germany\\
$^{94}$ Physikalisches Institut, Ruprecht-Karls-Universit\"{a}t Heidelberg, Heidelberg, Germany\\
$^{95}$ Physik Department, Technische Universit\"{a}t M\"{u}nchen, Munich, Germany\\
$^{96}$ Politecnico di Bari and Sezione INFN, Bari, Italy\\
$^{97}$ Research Division and ExtreMe Matter Institute EMMI, GSI Helmholtzzentrum f\"ur Schwerionenforschung GmbH, Darmstadt, Germany\\
$^{98}$ Saga University, Saga, Japan\\
$^{99}$ Saha Institute of Nuclear Physics, Homi Bhabha National Institute, Kolkata, India\\
$^{100}$ School of Physics and Astronomy, University of Birmingham, Birmingham, United Kingdom\\
$^{101}$ Secci\'{o}n F\'{\i}sica, Departamento de Ciencias, Pontificia Universidad Cat\'{o}lica del Per\'{u}, Lima, Peru\\
$^{102}$ Stefan Meyer Institut f\"{u}r Subatomare Physik (SMI), Vienna, Austria\\
$^{103}$ SUBATECH, IMT Atlantique, Nantes Universit\'{e}, CNRS-IN2P3, Nantes, France\\
$^{104}$ Sungkyunkwan University, Suwon City, Republic of Korea\\
$^{105}$ Suranaree University of Technology, Nakhon Ratchasima, Thailand\\
$^{106}$ Technical University of Ko\v{s}ice, Ko\v{s}ice, Slovak Republic\\
$^{107}$ The Henryk Niewodniczanski Institute of Nuclear Physics, Polish Academy of Sciences, Cracow, Poland\\
$^{108}$ The University of Texas at Austin, Austin, Texas, United States\\
$^{109}$ Universidad Aut\'{o}noma de Sinaloa, Culiac\'{a}n, Mexico\\
$^{110}$ Universidade de S\~{a}o Paulo (USP), S\~{a}o Paulo, Brazil\\
$^{111}$ Universidade Estadual de Campinas (UNICAMP), Campinas, Brazil\\
$^{112}$ Universidade Federal do ABC, Santo Andre, Brazil\\
$^{113}$ University of Cape Town, Cape Town, South Africa\\
$^{114}$ University of Houston, Houston, Texas, United States\\
$^{115}$ University of Jyv\"{a}skyl\"{a}, Jyv\"{a}skyl\"{a}, Finland\\
$^{116}$ University of Kansas, Lawrence, Kansas, United States\\
$^{117}$ University of Liverpool, Liverpool, United Kingdom\\
$^{118}$ University of Science and Technology of China, Hefei, China\\
$^{119}$ University of South-Eastern Norway, Kongsberg, Norway\\
$^{120}$ University of Tennessee, Knoxville, Tennessee, United States\\
$^{121}$ University of the Witwatersrand, Johannesburg, South Africa\\
$^{122}$ University of Tokyo, Tokyo, Japan\\
$^{123}$ University of Tsukuba, Tsukuba, Japan\\
$^{124}$ University Politehnica of Bucharest, Bucharest, Romania\\
$^{125}$ Universit\'{e} Clermont Auvergne, CNRS/IN2P3, LPC, Clermont-Ferrand, France\\
$^{126}$ Universit\'{e} de Lyon, CNRS/IN2P3, Institut de Physique des 2 Infinis de Lyon, Lyon, France\\
$^{127}$ Universit\'{e} de Strasbourg, CNRS, IPHC UMR 7178, F-67000 Strasbourg, France, Strasbourg, France\\
$^{128}$ Universit\'{e} Paris-Saclay Centre d'Etudes de Saclay (CEA), IRFU, D\'{e}partment de Physique Nucl\'{e}aire (DPhN), Saclay, France\\
$^{129}$ Universit\`{a} degli Studi di Foggia, Foggia, Italy\\
$^{130}$ Universit\`{a} del Piemonte Orientale, Vercelli, Italy\\
$^{131}$ Universit\`{a} di Brescia, Brescia, Italy\\
$^{132}$ Variable Energy Cyclotron Centre, Homi Bhabha National Institute, Kolkata, India\\
$^{133}$ Warsaw University of Technology, Warsaw, Poland\\
$^{134}$ Wayne State University, Detroit, Michigan, United States\\
$^{135}$ Westf\"{a}lische Wilhelms-Universit\"{a}t M\"{u}nster, Institut f\"{u}r Kernphysik, M\"{u}nster, Germany\\
$^{136}$ Wigner Research Centre for Physics, Budapest, Hungary\\
$^{137}$ Yale University, New Haven, Connecticut, United States\\
$^{138}$ Yonsei University, Seoul, Republic of Korea\\
$^{139}$  Zentrum  f\"{u}r Technologie und Transfer (ZTT), Worms, Germany\\
$^{140}$ Affiliated with an institute covered by a cooperation agreement with CERN\\
$^{141}$ Affiliated with an international laboratory covered by a cooperation agreement with CERN.\\

\end{flushleft} 

\end{document}